\documentclass[a4paper,11pt]{article}
\usepackage{pos}

\usepackage{nicefrac}
\usepackage{braket}
\usepackage{dsfont}
\usepackage{cleveref}

\usepackage{todonotes}

\title{A novel gauge-equivariant neural-network architecture for preconditioners in lattice QCD}
\ShortTitle{A novel GENN architecture for preconditioners in lattice QCD}

\author*{Simon Pfahler}
\author{Daniel Knüttel}
\author{Christoph Lehner}
\author{Tilo Wettig}

\affiliation{University of Regensburg, Department of Physics,\\
  93040 Regensburg, Germany}

\emailAdd{simon.pfahler@ur.de}

\abstract{
    Lattice QCD simulations are computationally expensive, with the solution of the Dirac equation being the major computational bottleneck of many calculations.
    We introduce a novel gauge-equivariant neural-network architecture for preconditioning the Dirac equation in the regime where critical slowing down occurs.
    We study the behavior of this preconditioner as a function of topological charge and lattice volume and show that it mitigates critical slowing down.
    We also show that this preconditioner transfers to unseen gauge configurations without any retraining, therefore enabling applications not possible with competing methods.
}

\FullConference{The 42nd International Symposium on Lattice Field Theory (LATTICE2025)\\
2-8 November 2025\\
Tata Institute of Fundamental Research, Mumbai, India\\}

\renewcommand{\logo}{\relax}

\begin{document}
\maketitle

\section{Introduction}
The main contribution to the cost of lattice QCD simulations typically comes from solving the Dirac equation.
Using iterative solvers, the iteration count required to reach convergence is determined by the condition number of the Dirac operator $D$.
In the limits of small lattice spacing and physical quark mass, the condition number and thus the iteration count diverges, a phenomenon known as critical slowing down.
A typical approach to mitigate this problem is the use of preconditioners.
A suitable preconditioner must address both the low and high modes of $D$ to reduce the condition number of the system.
The most effective preconditioner for this task is currently adaptive algebraic multigrid~\cite{Brannick_2008,Babich_2010,Frommer_2014}, which projects the low modes onto a coarser lattice where they can be addressed more easily.
Here, we investigate a different construction based on gauge-equivariant neural networks~\cite{Favoni_2022}.
By adapting the network architecture to allow for easier information transfer across long distances, we enable these models to also reduce the low-mode contributions to the residual vector.

A significant current limitation of algebraic multigrid is its setup cost, which consists of a few (typically $10-20$) inexact solves of the Dirac equation and is necessary for each gauge configuration.
We show that our neural-network preconditioners generalize to unseen gauge configurations without retraining, and that preconditioners trained on a smaller lattice can be successfully applied on a larger lattice without retraining.
This allows our method to be applied to cases in which algebraic multigrid's setup cost cannot be amortized easily, most prominently gauge-field generation.

\section{Previous Work}
Applying machine-learning methods to problems in lattice quantum field theories has seen a renewed interest in recent years~\cite{Kanwar_2020,Albergo_2021,Abbott_2022,Cali_2023,Sun_2025}.
Gauge-equivariant neural networks were introduced in this context in~\cite{Favoni_2022}, demonstrating that any gauge-equivariant function on the lattice can be approximated by a neural network.
It has already been shown that such networks can reduce the high-mode contributions to the residual vector effectively~\cite{Lehner_2023,Knuttel_2024}.
Alternative machine-learning-based approaches to training preconditioners for the Dirac equation have been used in~\cite{Cali_2023,Sun_2025} with some success.
So far, most of the work focused on addressing the high modes of the Dirac operator, apart from an exploratory study~\cite{Lehner_2024} that formulated multigrid in a machine-learnable way, but did not reduce the setup cost.
In this work, we extend and build upon~\cite{Lehner_2023,Knuttel_2024} to address also the low modes in order to build preconditioners that can mitigate critical slowing down.

\section{Theory}
\subsection{Problem statement}
The goal of this paper is to construct effective preconditioners for the Dirac equation
\begin{align}
	Du=b\,,
\end{align}
where in our context $D$ is a Wilson-clover Dirac operator for a given $\textrm{SU}(3)$ gauge field $U$, and $u$ and $b$ are quark fields.
When $D$ has small eigenvalues, its condition number diverges and the Dirac equation becomes ill-conditioned.
(Left) preconditioners $M$ are employed to counteract this effect by rewriting the original Dirac equation in the form
\begin{align}\label{eq:prec_Dirac_eq}
	MDu=Mb\,.
\end{align}
This equation has a more favorable condition number if $M$ approximates the inverse of $D$ sufficiently well, in particular for the lowest and highest eigenvalues of $D$.
There are many different ways to construct suitable preconditioners~\cite{Pearson_2020}.
In this work, we construct preconditioners using gauge-equivariant neural networks~\cite{Favoni_2022,Lehner_2023}.

\subsection{Network architecture}
The defining characteristic of a gauge-equivariant neural network is that its action commutes with gauge transformations.
The basic building block we use is the action of the hop operator (in space-time direction $\mu=\pm1,\ldots,\pm4$) on a quark field $\varphi$, given by\footnote{For negative directions $-\mu$, we identify $U_{-\mu}^\dag(x+\hat\mu)=U_\mu(x)$.}
\begin{align}
	(H_\mu\varphi)(x)=U_\mu^\dag(x-\hat\mu)\varphi(x-\hat\mu)\,.
\end{align}
Note that $H_\mu\varphi$ is the resulting field, evaluated at space-time position $x$.
It can easily be checked that this operation is gauge-equivariant~\cite{Lehner_2023}.
We also define $H_0=\mathds1$.
By concatenating hops $H_\mu$, we can define the parallel-transport operator $T_p$ for a path $p=p_1,\ldots,p_{n_p}$ consisting of single hops $p_i=\pm\mu$ as
\begin{align}
	T_p=H_{p_{n_p}}\cdots H_{p_1}\,.
\end{align}
We introduce a parallel-transport layer (PT) that takes as an input $n$ quark fields $\varphi_a$ and applies a parallel-transport operator to each of them (using an $a$-dependent path) to construct $n$ output fields $\psi_a$,
\begin{align}
	\psi_a(x)\overset{\textrm{PT}}=T_{p_a}\varphi_a(x)\,,\qquad a=1,\ldots,n\,.
\end{align}
Additionally, we introduce a linear layer (L) that takes $n$ quark fields and constructs $m$ linear combinations from them,
\begin{align}
	\psi_b(x)\overset{\textrm{L}}=\sum_{a=1}^nW_{ba}\varphi_a(x)\,,\qquad b=1,\ldots,m\,.
\end{align}
The parameters $W_{ba}\in\mathbb C^{4\times4}$ are learnable weights of this layer that act on the spinor degrees of freedom of the quark fields.

\begin{figure}
	\centering
	\includegraphics{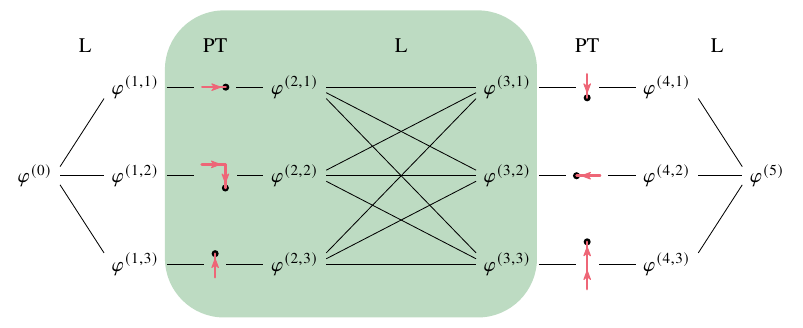}
	\caption{Exemplary network architecture using linear layers (L) and parallel-transport layers (PT).
		For deeper networks, the highlighted block consisting of a PT and an L layer can be repeated, with potentially different parallel-transport paths in each of the PT layers.}
	\label{fig:network_architecture}
\end{figure}

By alternating between L and PT layers, we can construct trainable neural networks.
Note that the PTC layer used in~\cite{Lehner_2023,Knuttel_2024} can be constructed from these building blocks, and hence our architecture can be seen as a generalization of the PTC layer.
In order to use such a neural network as a preconditioner, the network may have only one input and output quark field, see \cref{eq:prec_Dirac_eq}.
The intermediate layers, however, may contain more fields.
Therefore, a suitable architecture for the neural network is to first employ a one-to-many L layer, followed by alternating PT and L layers, with the final L layer reducing the number of fields back to one.
\Cref{fig:network_architecture} shows an exemplary network architecture.

Apart from the basic network architecture, the choice of parallel-transport paths in the network also has a large influence on the resulting iteration-count reduction.
In this work, we consider two different choices of parallel-transport paths.
The first choice is to simply use single forward and backward hops in all space-time dimensions, as well as no hop, resulting in the set of paths
\begin{align}\label{eq:Ps}
	P_s=\{0;\pm1;\pm2;\pm3;\pm4\}\,.
\end{align}
This is a natural choice of paths, but it has the drawback that transporting information over a distance of $d$ lattice sites requires $d$ such layers.
If we want the neural network to connect all sites on a lattice of size $L_1\times L_2\times L_3\times L_4$, we therefore need at least $\mathcal O(L)$ layers, where $L=\max\{L_\mu\}$.

As a second choice of parallel-transport paths, we add more straight paths that increase by factors of $2$ in length.
Using the notation $(\mu, \ell_\mu)$ for a straight path in direction $\mu$ covering a distance $\ell_\mu$, we consider the set of paths
\begin{align}\label{eq:Pl}
	P_\ell=\{0\}\cup\big\{(\mu,\ell_\mu)\;\big|\;\mu\in\{\pm1,\pm2,\pm3,\pm4\},\,\ell_\mu\in\{2^0,2^1,2^2,\ldots,\nicefrac{L_\mu}2\}\big\}\,.
\end{align}
By incorporating longer paths, the number of layers needed to connect all lattice sites is drastically reduced to $\mathcal O(\log(L))$.
Since we are limiting ourselves to straight paths, the number of paths in each layer remains manageable.

\subsection{Filtered cost function}
So far we defined the network architecture of our preconditioner.
To train the network weights via machine learning, we need to define a differentiable cost function that our network should minimize.
The natural choice for this would be the condition number
\begin{align}
	\kappa=\frac{|\lambda_\textrm{max}|}{|\lambda_\textrm{min}|}\,.
\end{align}
In order to use the condition number as a cost function, the eigenvalues would need to be computed while keeping track of gradients via automatic differentiation.
However, this process is computationally very expensive as the eigenvalues of the Dirac operator are not readily available.
Additionally, performing automatic differentiation through this calculation is numerically unstable~\cite{Sun_2025,Hovland_2024}.
The condition number is therefore not a viable cost function in our case.

Prior works have therefore resorted to using the surrogate cost function
\begin{align}
	C=\|MDv-v\|^2
\end{align}
with random training vectors $v$~\cite{Lehner_2023,Sun_2025}.
While this cost function has the desired minimum at $M=D^{-1}$ and can easily be calculated, it emphasizes the large eigenmodes of $D$.
To see this, we write the preconditioner as $M=D^{-1}+E$ with an error matrix $E$ that should be minimized by a well-trained network.
The matrix $D$ is non-Hermitian and has a right and a left eigensystem given by
\begin{align}
	D\ket{r_i}=\lambda_i\ket{r_i},\quad\bra{\ell_i}D=\lambda_i\bra{\ell_i}\,,
\end{align}
where the $\ket{r_i}$ are normalized to unity and the $\bra{\ell_i}$ are normalized such that $\braket{\ell_i|r_j}=\delta_{ij}$.
We also have $\sum_i\ket{r_i}\bra{\ell_i}=\mathds{1}=\sum_i\ket{\ell_i}\bra{r_i}$.
This allows us to rewrite the cost function as
\begin{align}
	C=\sum_{ij}\lambda_i^*\lambda_jv_i^*v_j(E^\dag E)_{ij}
\end{align}
with $v_i=\braket{\ell_i|v}$ and $(E^\dag E)_{ij}=\braket{r_i|E^\dag E|r_j}$.
Crucially, when we have near-zero eigenvalues, the corresponding eigenmodes contribute very little to the cost function and the network is not incentivized to minimize the corresponding entries in $E^\dag E$.

To counteract this, we introduce the filtered cost function
\begin{align}\label{eq:filtered_cost}
	C_N=\|MDu_N-u_N\|^2\,,
\end{align}
where $u_N$ is an approximate solution of the linear system $D^ku_N=0$ obtained from $N$ iterations of GMRES with a random initial guess.
Writing $u_N=\sum_iu_i\ket{r_i}$ we have $D^ku_N=\sum_i\lambda_i^ku_i\ket{r_i}\approx0$.
Since the $\ket{r_i}$ are linearly independent, the combination $\lambda_i^ku_i$ should be small for all $i$, i.e., the $u_i$ should behave like $\varepsilon_i/\lambda_i^k$ with small $\varepsilon_i$ so that
\begin{align}
	C_N=\sum_{ij}(\lambda_i^*\lambda_j)^{1-k}\varepsilon_i^*\varepsilon_j(E^\dagger E)_{ij}\,.
\end{align}
The choice of $k$ allows us to shift the focus of the training between the high and low modes.
In our tests below, we set $k=1$ to treat the high and low modes equally.

\section{Numerical experiments}
When testing the benefit of our networks as preconditioners, we always consider residuals achieved after a given number of operator applications, not simply iterations of the solver.
We define an operator application as an application of the Dirac operator or an application of a PT layer in our model, as these operations are formally very similar and dominate the runtime of the solver.
The cost of an L layer is negligible and therefore omitted.
Considering iteration count directly would omit the cost of employing the preconditioner and would produce misleading results.

\subsection{Choice of parallel-transport paths and cost function}
In all of our numerical experiments, we use quenched gauge configurations with the Wilson gauge action at $\beta=6$.

\begin{figure}
	\centering
	\hfill\includegraphics{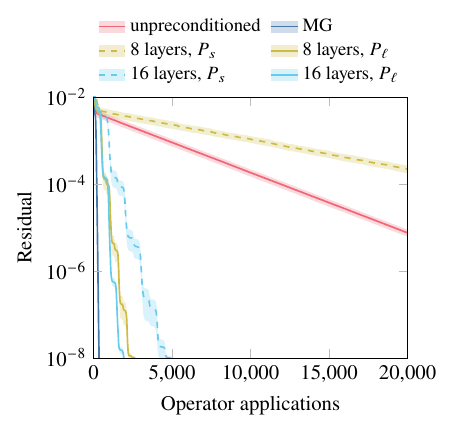}
	\includegraphics{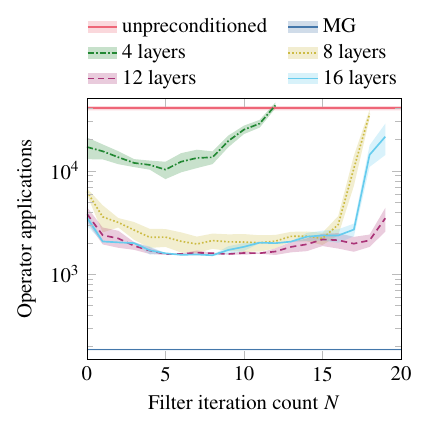}\hfill
	\caption{Left: Evolution of the residual during GMRES solves with and without preconditioners, comparing the choices $P_s$ and $P_\ell$ of parallel-transport paths.
		Networks are trained with the cost function $C_N$ with $N=10$ filter iterations.
		Right: Operator applications needed in GMRES solves to reach a residual of $10^{-18}$ with and without preconditioners, where all neural networks use the set $P_\ell$ of parallel-transport paths.
		Both plots show data for a lattice size of $8^3\times 16$, a quenched gauge configuration at $\beta=6$ with topological charge $Q=1$, and bare mass parameter $m=-0.555$ (such that $m-m_\textrm{crit}\approx5\times10^{-4}$).}
	\label{fig:paths_and_filter}
\end{figure}

First, we test the effect of the different choices of parallel-transport paths in \cref{eq:Ps,eq:Pl} and the number $N$ of filter iterations in the cost function $C_N$ in \cref{eq:filtered_cost}.
To compare the choices of parallel-transport paths, we train networks with different numbers of PT layers for both $P_s$ and $P_\ell$ using the cost function \cref{eq:filtered_cost} with $N=10$ iterations of the filter.
After the training process converged\footnote{Convergence was obtained after $1000$ training steps.}, we can use the networks as preconditioners for GMRES in test solves of the Dirac equation.
To assess the speedup of the solver, we consider the evolution of the residual during the solving process.
The left plot in \cref{fig:paths_and_filter} shows the performance of models with both choices of paths and compares them to the unpreconditioned and the multigrid-preconditioned solve.
The step-like behavior that can be observed in the plot is due to the restart length in GMRES, which is chosen such that GMRES builds a new Krylov space every $30$ iterations.
It can be seen that both prescriptions for the parallel-transport paths ($P_s$ and $P_\ell$) are beneficial as a preconditioner, as long as a sufficient number of layers is used.
The plot also demonstrates that choosing the set $P_\ell$ leads to more effective preconditioners than choosing $P_s$.
This behavior was observed for different gauge fields, masses, lattice sizes, and filter iterations.
For the remainder of this work, we therefore limit ourselves to the choice $P_\ell$.

In the right plot of \cref{fig:paths_and_filter}, we can see the number of operator applications needed for various GMRES solves to reach a residual of $10^{-8}$, depending on the number $N$ of filter iterations in $C_N$ used in the training process.
We can see that a nonzero choice of $N$ is beneficial for model performance as it decreases the number of operator applications by as much as a factor of $3$ in many cases, and that model performance is not very sensitive to the precise choice of $N$.
Therefore, we fix $N=10$ for the remainder of this work.

\subsection{Solver speedup}
Now that we have found suitable choices of parallel-transport paths and the cost function, we can train preconditioners for different lattice sizes, topological charges, and bare mass parameters and check how our models manage to deal with critical slowing down.

\begin{figure}
	\centering
	\hfill\includegraphics[page=1]{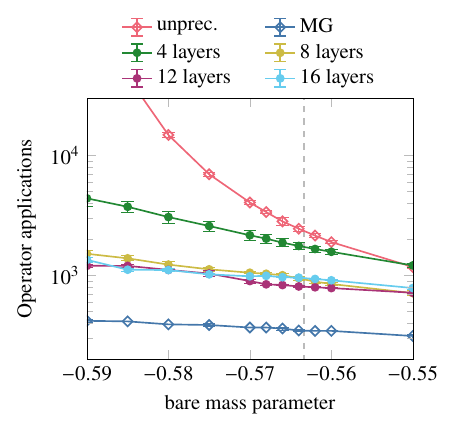}
	\includegraphics[page=2]{graphics/csd.pdf}\hfill
	\caption{Operator applications needed in GMRES solves to reach a residual of $10^{-18}$ with and without preconditioners, depending on the bare mass parameter.
		An $8^3\times16$ lattice is used, and the topological charge of the gauge configuration is $Q=0$ (left) and $Q=1$ (right).
		Models are trained individually for each bare mass parameter.
		The dashed vertical line denotes the critical mass, defined as the largest bare mass parameter for which an eigenvalue of $D$ has zero real part.
	}
	\label{fig:CSD_8c16}
\end{figure}

\Cref{fig:CSD_8c16} shows results for on an $8^3\times16$ lattice for topological charges $Q=0$ and $Q=1$.
For both topological charges, the trained networks outperform the unpreconditioned solve significantly, and they exhibit more favorable scaling with respect to the bare mass parameter.
Near criticality, our preconditioners reduce the required number of operator applications by over an order of magnitude for $Q=1$.
However, we see that critical slowing down is not completely avoided, as the operator applications needed for all trained models still exhibit a divergence when approaching the critical mass, even though it is greatly reduced compared to the unpreconditioned case.

\begin{figure}
	\centering
	\hfill\includegraphics[page=5]{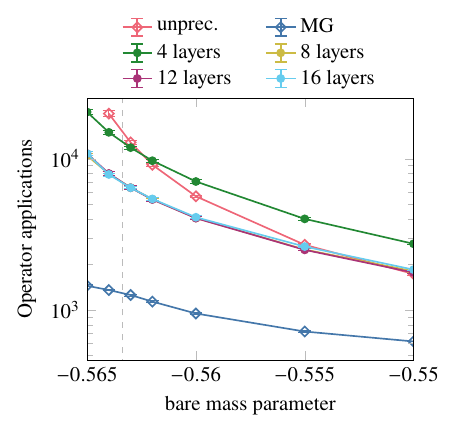}
	\includegraphics[page=6]{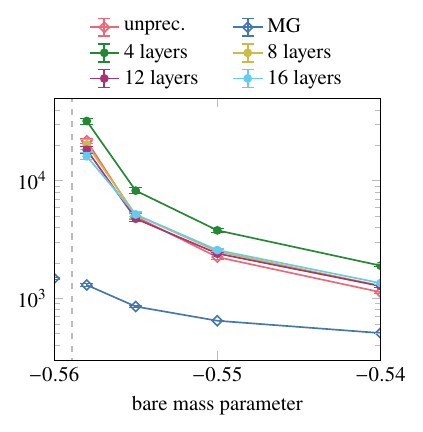}\hfill
	\caption{Same as \cref{fig:CSD_8c16}, but for a $16^3\times32$ lattice and topological charges $Q=0$ (left) and $Q=4$ (right).}
	\label{fig:CSD_16c32}
\end{figure}

We repeat this test on a $16^3\times32$ lattice, see \cref{fig:CSD_16c32}.
For $Q=0$, we see that our preconditioners still exhibit favorable scaling when compared to the unpreconditioned solve, but the benefit is rather small in this case.
Also, our model does not achieve the performance of multigrid.
This effect is even stronger for topological charge $Q=4$, where the benefit of using our networks as preconditioners is marginal if at all present.
This behavior suggests that our networks currently have a problem with addressing topological modes properly.

\subsection{Transfer to unseen gauge configurations}
\begin{figure}
	\centering
	\hfill\includegraphics[page=2]{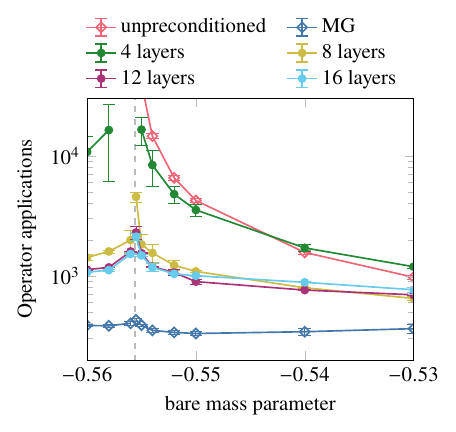}
	\includegraphics[page=6]{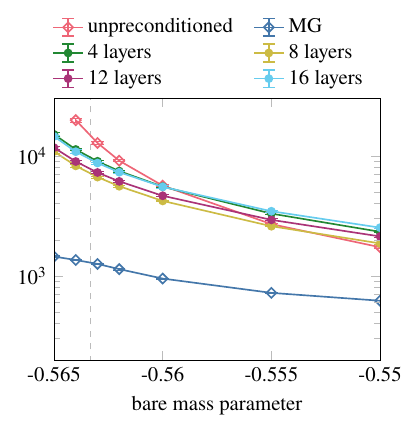}\hfill
	\caption{Application of a model trained on the $8^3\times16$ lattice with a gauge configuration with $Q=0$ and $m=-0.56$ to a $8^3\times16$ lattice with a gauge configuration with $Q=1$ and different masses (left), and to a $16^3\times32$ lattice with a gauge configuration with $Q=0$ and different masses (right).}
	\label{fig:transfer}
\end{figure}

A major goal of this research project is to construct preconditioners that are competitive with multigrid for each solve, but have a smaller setup cost.
In particular, we want to create networks that, once trained, can be used to precondition a Dirac operator for any gauge field without requiring additional training on the specific configuration.
\Cref{fig:transfer} shows this transfer of networks which were trained on an $8^3\times16$ lattice for a configuration with topological charge $Q=0$ and bare mass parameter $m=-0.56$.
The left plot shows the application of these models to an $8^3\times16$ lattice and a configuration with topological charge $Q=1$ (compare to the right plot of \cref{fig:CSD_8c16}, which shows the performance without transfer), and the right plot shows the application to a $16^3\times32$ lattice and a configuration with $Q=0$ (compare to the left plot in \cref{fig:CSD_16c32}).
From the comparison, we see in both cases that transfer works very well, as the transferred network performs on average as well as the network trained for the configuration.
This shows that our approach has the potential of leading to preconditioners with drastically lower setup cost than multigrid while still addressing critical slowing down.

\section{Summary and outlook}
We proposed a novel gauge-equivariant neural-network architecture that aims to address the critical slowing down of iterative solvers in lattice QCD observed when approaching the physical limits.
We generalized the gauge-equivariant building block introduced in~\cite{Lehner_2023} to allow for more expressive networks without increasing the number of layers.
Additionally, we introduced a filter into the cost function typically used to train such models, which allows us to focus more strongly on the low modes of the operator and thus the problematic part of the spectrum when dealing with critical slowing down.
In numerical experiments, we showed that our preconditioning networks provide a significant benefit in time-to-solution, and in particular that employing them leads to a more beneficial runtime scaling when moving closer to the critical mass.
Importantly, our trained models transfer to unseen gauge configurations, regardless of their topological charge or their lattice volume.

In future work, we need to address the remaining challenges that arise when going to large topological charge and larger lattice volumes.
While results are very promising on a small $8^3\times16$ lattice, application to $16^3\times32$ lattices fails to show the same level of solver speedup.
In order to address this, we will analytically investigate the trained weights to find out if the models uncover a general structure, as this seems likely due to the perfect transferability of models.
Additionally, by considering simpler toy models, we hope to understand why our models have problems for large lattice sizes and nonzero topological charge.
Isolating the problematic behavior in this way will enable us to understand and hopefully solve the remaining challenges towards setup-free preconditioners that can compete with multigrid.

\bibliographystyle{JHEP}
\bibliography{references_LatticePoS}

\providecommand{\href}[2]{#2}\begingroup\raggedright\begin{thebibliography}{10}

\bibitem{Brannick_2008}
J.~Brannick, R.C.~Brower, M.A.~Clark, J.C.~Osborn and C.~Rebbi, \emph{Adaptive {{Multigrid Algorithm}} for {{Lattice QCD}}}, \href{https://doi.org/10.1103/PhysRevLett.100.041601}{\emph{Physical Review Letters} {\bfseries 100} (2008) 041601}.

\bibitem{Babich_2010}
R.~Babich, J.~Brannick, R.C.~Brower, M.A.~Clark, T.A.~Manteuffel, S.F.~McCormick et~al., \emph{Adaptive multigrid algorithm for the lattice {{Wilson-Dirac}} operator}, \href{https://doi.org/10.1103/PhysRevLett.105.201602}{\emph{Physical Review Letters} {\bfseries 105} (2010) 201602} [\href{https://arxiv.org/abs/1005.3043}{{\ttfamily 1005.3043}}].

\bibitem{Frommer_2014}
A.~Frommer, K.~Kahl, S.~Krieg, B.~Leder and M.~Rottmann, \emph{Adaptive {{Aggregation-Based Domain Decomposition Multigrid}} for the {{Lattice Wilson--Dirac Operator}}}, \href{https://doi.org/10.1137/130919507}{\emph{SIAM Journal on Scientific Computing} {\bfseries 36} (2014) A1581}.

\bibitem{Favoni_2022}
M.~Favoni, A.~Ipp, D.I.~M{\"u}ller and D.~Schuh, \emph{Lattice {{Gauge Equivariant Convolutional Neural Networks}}}, \href{https://doi.org/10.1103/PhysRevLett.128.032003}{\emph{Physical Review Letters} {\bfseries 128} (2022) 032003}.

\bibitem{Kanwar_2020}
G.~Kanwar, M.S.~Albergo, D.~Boyda, K.~Cranmer, D.C.~Hackett, S.~Racani{\`e}re et~al., \emph{Equivariant {{Flow-Based Sampling}} for {{Lattice Gauge Theory}}}, \href{https://doi.org/10.1103/PhysRevLett.125.121601}{\emph{Physical Review Letters} {\bfseries 125} (2020) 121601}.

\bibitem{Albergo_2021}
M.S.~Albergo, D.~Boyda, D.C.~Hackett, G.~Kanwar, K.~Cranmer, S.~Racani{\`e}re et~al., \emph{Introduction to {{Normalizing Flows}} for {{Lattice Field Theory}}},  Aug., 2021.
\newblock 10.48550/arXiv.2101.08176.

\bibitem{Abbott_2022}
R.~Abbott, M.S.~Albergo, D.~Boyda, K.~Cranmer, D.C.~Hackett, G.~Kanwar et~al., \emph{Gauge-equivariant flow models for sampling in lattice field theories with pseudofermions}, \href{https://doi.org/10.1103/PhysRevD.106.074506}{\emph{Physical Review D} {\bfseries 106} (2022) 074506}.

\bibitem{Cali_2023}
S.~Cal{\`i}, D.C.~Hackett, Y.~Lin, P.E.~Shanahan and B.~Xiao, \emph{Neural-network preconditioners for solving the {{Dirac}} equation in lattice gauge theory}, \href{https://doi.org/10.1103/PhysRevD.107.034508}{\emph{Physical Review D} {\bfseries 107} (2023) 034508}.

\bibitem{Sun_2025}
Y.~Sun, S.~Eswar, Y.~Lin, W.~Detmold, P.~Shanahan, X.~Li et~al., \emph{Matrix-free {{Neural Preconditioner}} for the {{Dirac Operator}} in {{Lattice Gauge Theory}}},  Sept., 2025.
\newblock 10.48550/arXiv.2509.10378.

\bibitem{Lehner_2023}
C.~Lehner and T.~Wettig, \emph{Gauge-equivariant neural networks as preconditioners in lattice {{QCD}}}, \href{https://doi.org/10.1103/PhysRevD.108.034503}{\emph{Physical Review D} {\bfseries 108} (2023) 034503}.

\bibitem{Knuttel_2024}
D.~Kn{\"u}ttel, C.~Lehner and T.~Wettig, \emph{Gauge-equivariant multigrid neural networks}, \href{https://doi.org/10.22323/1.453.0037}{\emph{PoS} {\bfseries LATTICE2023} (2024) 037}.

\bibitem{Lehner_2024}
C.~Lehner and T.~Wettig, \emph{Gauge-equivariant pooling layers for preconditioners in lattice {{QCD}}}, \href{https://doi.org/10.1103/PhysRevD.110.034517}{\emph{Physical Review D} {\bfseries 110} (2024) 034517}.

\bibitem{Pearson_2020}
J.W.~Pearson and J.~Pestana, \emph{Preconditioners for {{Krylov}} subspace methods: {{An}} overview}, \href{https://doi.org/10.1002/gamm.202000015}{\emph{GAMM-Mitteilungen} {\bfseries 43} (2020) e202000015}.

\bibitem{Hovland_2024}
P.~Hovland and J.~H{\"u}ckelheim, \emph{Differentiating {{Through Linear Solvers}}},  May, 2024.
\newblock 10.48550/arXiv.2404.17039.

\end{thebibliography}\endgroup

\end{document}